\documentclass[aps, prd, twocolumn, superscriptaddress, nofootinbib, noshowpacs, preprintnumbers, longbibliography, floatfix]{revtex4-2}

\usepackage[utf8]{inputenc}
\usepackage[T1]{fontenc}
\usepackage{lmodern}
\usepackage{indentfirst}
\usepackage{graphicx}
\usepackage{rotating}
\usepackage{placeins}
\usepackage{float}
\usepackage{amsmath, amssymb, amsfonts}
\usepackage{amsthm}
    {
    \theoremstyle{definition}
    \newtheorem{asmp}{Assumption}[section]
    }
\usepackage{mathtools}
\usepackage{adjustbox}
\usepackage{diagbox}
\usepackage{array}
\usepackage{multirow}
\usepackage[ruled,vlined]{algorithm2e}
\usepackage{xcolor}
\usepackage{physics}
\usepackage{tikz}
\usetikzlibrary{quantikz2}
\usepackage{subfigure}
\usepackage{braket}
\usepackage{slashed}
\usepackage[colorlinks,
    citecolor={blue!70!black},
    urlcolor={blue!70!black},
    linkcolor={red!70!black},
    hyperindex,breaklinks]{hyperref}
\usepackage{orcidlink}

\usepackage{soul}

\newcommand{\FirstAffiliation}{\affiliation{
    Universit\'{e} Paris-Saclay,
    CEA,
    List,
    F-91120,
    Palaiseau,
    France
    }}
    
\begin{document}
%
%
%


\title{Phase Diagram of the Schwinger Model by Adiabatic Preparation of States on a Quantum Simulator}



\author{Oleg Kaikov \orcidlink{0000-0002-9473-7294}}
\email{oleg.kaikov@cea.fr}
\FirstAffiliation

\author{Theo Saporiti \orcidlink{0009-0008-9738-3402}}
\FirstAffiliation

\author{Vasily Sazonov \orcidlink{0000-0002-8152-0221}}
\FirstAffiliation

\author{Mohamed Tamaazousti \orcidlink{0000-0002-3947-9069}}
\FirstAffiliation


\date{\today}


\begin{abstract}
\noindent
We argue the feasibility to study the phase structure of a quantum physical system on quantum devices via adiabatic preparation of states. We introduce a novel method and successfully test it in application to the Schwinger model in the presence of a topological $\theta$-term. We explore the first-order-phase-transition and the no-transition regions of the corresponding phase diagram. The core idea of the method is to separately evolve the ground and the first excited states with a time-dependent Hamiltonian, the time-dependence of which interpolates between different values of $\theta$. Despite our approach being a direct application of the adiabatic theorem, in some cases we are able to demonstrate its advantages in comparison to a different method from the literature that also employs adiabatic state preparation.
\end{abstract}

\maketitle


\section{Introduction}

Within any physical theory, the phase diagram determines the behavior of the fields for any given parameter regime. The knowledge of the phase structure of a theory is therefore fundamental for the understanding of the physical phenomena that it models. Due to the complexity of many theories, their phase diagrams are often studied numerically. An approach commonly utilized for this task is that of Monte Carlo (MC) simulation. For example, this method has been successfully used to analyze certain parameter regimes within the phase diagram of lattice Quantum Chromodynamics (QCD)~\cite{10.1143/PTPS.174.206}.

However, for lattice QCD (LQCD) this approach fails in the presence of a non-zero chemical potential~\cite{deForcrand:2010j3, NAGATA2022103991}, resulting in a non-zero baryon density: The Boltzmann factor of the Euclidean path integral becomes complex. This factor can therefore no longer be reliably considered as a probability weight in MC algorithms. Such an occurrence of a negative or complex probability measure is referred to as the sign problem. In addition, a non-zero charge conjugation and parity symmetry (CP) violating topological $\theta$-term in LQCD would also lead to the sign problem~\cite{CarmenBañuls_2020}. Thus, adhering to the MC method requires modifications~\cite{10.1143/PTPS.174.206}. However, other promising approaches~\cite{CarmenBañuls_2020, Bañuls2020, Funcke:2023RU, halimeh2023coldatom}, that are unaffected by the sign problem, are those of tensor networks and quantum computing. For example, considerable progress has already been achieved in the latter direction~\cite{Martinez2016, Yang2020, PRXQuantum.2.030334, BassmanOftelie_2021, Thompson_2022, PRXQuantum.3.020324, zhang2023observation, dimeglio2023quantum, farrell2024quantum, PRXQuantum.5.020315}.

In this work we argue the feasibility to explore the phase diagram of the Schwinger model~\cite{PhysRev.128.2425} with a non-zero topological $\theta$-term by adiabatic preparation of states on a digitally simulated quantum device. Despite the fact that the Schwinger model is less expressive than QCD, the two models share many key properties. Therefore, on the one hand, the Schwinger model is easier to study. On the other hand, due to its similarity to QCD, the Schwinger model serves as a good benchmark theory for testing various methods designed to be applied to QCD in the future.

Within the conventional lattice formulation, the presence of a topological $\theta$-term in the Schwinger model renders the standard MC methods inapplicable due to the sign problem. Many analytic solutions have been developed in the literature. For example, there are methods based on a reformulation of the theory, allowing the use of the standard MC approach. These include, e.g., a dual formulation within the massless lattice Schwinger model with staggered fermions~\cite{GATTRINGER2015732, GOSCHL201763}, and the lattice formulation of the bosonized Schwinger model~\cite{BENDER1985745, Ohata2023, 10.1093/ptep/ptad151} for arbitrary fermion mass. In the present work, however, we explore the quantum computing approach.

The Schwinger model with a non-zero $\theta$-term possesses a diverse phase structure including a first-order phase transition (PT) at $\theta = \pi$ for sufficiently large masses and a region without transitions in the sub-critical mass regime~\cite{COLEMAN1975267, COLEMAN1976239, HAMER1982413, PhysRevD.66.013002, PhysRevD.97.014507, PhysRevD.101.054507, PhysRevD.105.014504, Pederiva:2022br, PhysRevD.108.014516, angelides2024firstorder}. This makes the Schwinger model the perfect candidate to test our approach utilizing a quantum computer. Although analyses of the Schwinger model and its multi-flavor versions include different variational~\cite{angelides2024firstorder, PhysRevLett.118.071601} and adiabatic~\cite{PhysRevD.105.014504, Pederiva:2022br, PhysRevA.90.042305, Yamamoto:2022Qn, PhysRevD.105.094503, ghim2024digital} methods, to the best of our knowledge, we are not aware of studies applying the method that we propose to this or any other model.

Variational approaches, such as the variational quantum eigensolver (VQE)~\cite{Peruzzo2014}, can be computationally cheaper in comparison to adiabatic methods for some particular problems. However, the choice of an appropriate variational ansatz always requires additional insight into the system being studied in order to exclude undesired sub-regions of the Hilbert space. Moreover, in general, the VQE optimization problem is NP-hard~\cite{PhysRevLett.127.120502}. Conversely, although adiabatic methods can be computationally more costly in specific cases, they are comparatively simpler to employ. Our method belongs to the latter category and is based on evolving multiple eigenstates with a time-dependent Hamiltonian, the time-dependence of which interpolates between different values of the considered parameter, which, in our case, is the value of $\theta$.

In section~\ref{sect:Schw_model} we briefly review the Schwinger model in the continuum, its phase structure and its lattice discretization via staggered fermions. In section~\ref{sect:adiab_st_prep} we provide an overview of several versions of the adiabatic theorem, define the necessary notation and present our method. In section~\ref{sect:num_an} we present the numerical results of our approach. In section~\ref{sect:disc} we discuss our method and compare it to a different method that utilizes adiabatic state preparation as well as a deformation of the Hamiltonian to avoid level crossings~\cite{sym14040809}. Last, in section~\ref{sect:concl} we present our conclusions and an outlook on future work.

\section{The Schwinger Model} \label{sect:Schw_model}

The Schwinger model is a (1+1)-dimensional $U(1)$ gauge theory. It describes quantum electrodynamics coupled to a single flavor of a massive Dirac fermion~\cite{PhysRev.128.2425}. In this section we review the Hamiltonian formulation of the theory and the phase diagram of the model.

\subsection{Continuum theory}

We set $\hbar = c = 1$ throughout the work. The continuum Hamiltonian density of the Schwinger model in the presence of the $\theta$-term reads
\begin{equation} \label{eqn:Ham_dens}
\mathcal{H} = -i \overline{\psi} \gamma^1 \left( \partial_1 -i g A_1 \right) \psi + m \overline{\psi} \psi + \dfrac{1}{2} \left( \dot{A}_1 + \dfrac{g \theta}{2 \pi} \right)^2 \, .
\end{equation}
Here, $\psi$ is a two-component Dirac spinor. The components $\psi_{\alpha}(x)$, $\alpha=1,2$, obey the usual fermionic anti-commutation relations $\{ \psi^{\dagger}_{\alpha}(x), \psi_{\beta}(y) \} = \delta(x-y)\delta_{\alpha \beta}$. The gauge field is denoted by $A_{\mu}$, $\mu=0,1$, with the Hamiltonian density (\ref{eqn:Ham_dens}) written in the temporal gauge $A_0 = 0$. The parameters $m$ and $g$ denote the bare fermion mass and coupling, respectively. The two-dimensional $\gamma^{\mu}$ matrices obey the Clifford algebra $\{\gamma^{\mu},\gamma^{\nu}\}=2\eta^{\mu\nu}$, where $\eta=\text{diag}(1,-1)$. We adopt the convention $\gamma^0=X$, $\gamma^1=iZ$, with the standard Pauli matrices $X$ and $Z$. We also identify $\overline{\psi}=\psi^{\dagger}\gamma^0$. The physical states of the theory satisfy Gauss's law
\begin{equation} \label{eqn:Gauss_cont}
- \partial_1 \dot{A}^1 = g \overline{\psi} \gamma^0 \psi \, .
\end{equation}

We now discuss the properties of the model. A non-zero $\theta$-term in (\ref{eqn:Ham_dens}) results in a constant background electric field. The physics of the model is periodic in $\theta$ with a period of $2\pi$. At $\theta = \pi$, there is a second-order PT at the critical mass in units of the coupling $m_{\text{c}}/g \approx 0.33$ and a first-order PT above $m_{\text{c}}/g$ at the same value of $\theta$~\cite{HAMER1982413, COLEMAN1976239, PhysRevD.66.013002, PhysRevD.95.094509}. Qualitatively, this can be understood as follows: For $m/g \gg 1$ and a background electric field $\theta < \pi$, it is energetically costly to produce charged particles. Therefore, in this regime the ground state corresponds to a state with no particles. Contrarily, for $m/g \gg 1$ and a background electric field $\theta > \pi$ it is energetically favorable to produce a negatively charged particle on the left of the one-dimensional interval and a positively charged particle on the right. Correspondingly, the electric field is reduced by one unit~\cite{COLEMAN1976239, PhysRevD.66.013002}. These two parameter regions are separated by a first-order PT line at $\theta = \pi$ for $m/g \gg 1$, where the two corresponding ground states are degenerate in energy.

\begin{sidewaysfigure*}[htbp]
    \centering
    \vspace{10cm}
    \includegraphics[width=\textwidth]{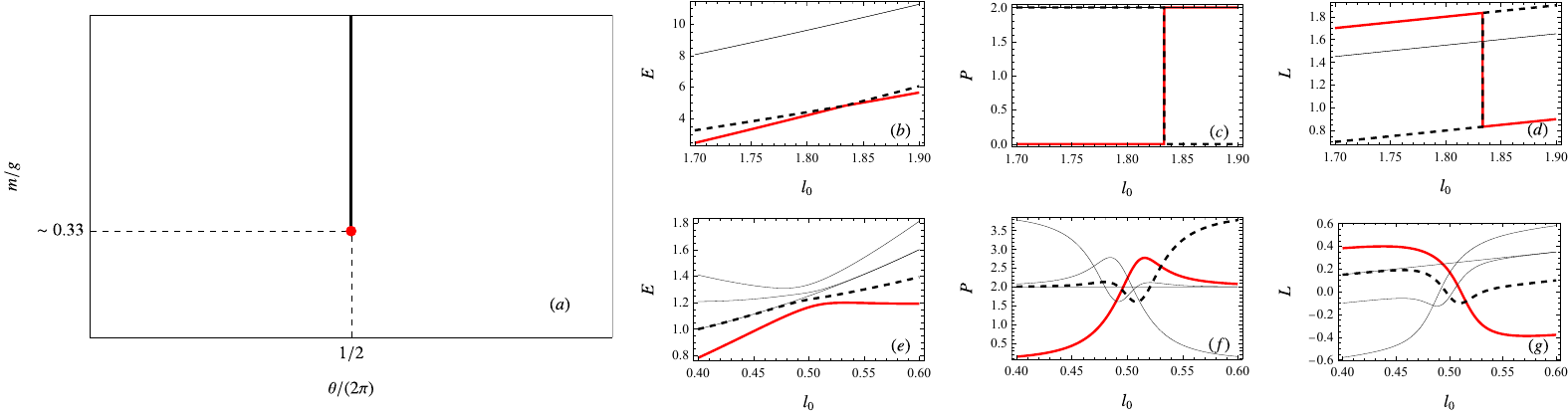}
    \caption{\small Phase structure of the Schwinger model. Panel (a): phase diagram of the model in the $\theta$ - $m/g$ plane. For $m/g > m_{\text{c}}/g \approx 0.33$ there is a first-order phase transition at $\theta = \pi$ (solid black line). This first-order phase transition line ends at $m_{\text{c}}/g$ in a second-order phase transition (red point). Below $m_{\text{c}}/g$ no transitions take place. Panels (b) - (g): exact diagonalization results (see section~\ref{sect:num_an} for numerical values of input parameters) for the eigenstate energy $E$ (panels (b), (e)), particle number $P$ (panels (c), (f)) and electric field density $L$ (panels (d), (g)) for $10^3$ steps over $l_0$ for a set of lowest eigenstates. The results for the ground state, first excited state and higher excited states are shown in solid red, dashed black and thin solid black, respectively. Panels (b) - (d): four lowest eigenstates for $m_{\text{lat}}/g = 10$ (above second-order phase transition). The second and third excited states (if not degenerate) cannot be distinguished at this scale. Panels (e) - (g): five lowest eigenstates for $m_{\text{lat}}/g = 0$ (below second-order phase transition).}
    \label{fig:phase_str}
\end{sidewaysfigure*}

Conversely, for $m/g < m_{\text{c}}/g$ it is energetically more favorable to produce more pairs of charged particles. This screens the background electric field~\cite{ADAM19971}. Therefore, there is no PT in this parameter region. The corresponding phase diagram of the Schwinger model is shown in Fig.~\ref{fig:phase_str}(a).

\subsection{Lattice formulation} \label{subsect:lat_form}

To study the Schwinger model numerically we utilize the lattice discretization via Kogut-Susskind staggered fermions~\cite{PhysRevD.11.395}, following the recent work~\cite{angelides2024firstorder}. Here we closely recapitulate the main steps of~\cite{angelides2024firstorder}. In this formulation, the Schwinger model Hamiltonian on a lattice with an even number of sites $N$ (we consider $N \geq 4$) with spacing $a$ and open boundary conditions (OBC) reads~\cite{PhysRevD.11.395}
\begin{equation} \label{eqn:Ham_stagg}
    \begin{aligned}
        H_{\text{lat}} &= -\dfrac{i}{2a} \sum\limits_{n=0}^{N-2} \left( \phi^{\dagger}_n U_n \phi_{n+1} - \text{H.c.} \right) \\
        &+ m_{\text{lat}} \sum\limits_{n=0}^{N-1} (-1)^n \phi^{\dagger}_n \phi_n \\
        &+ \dfrac{a g^2}{2} \sum\limits_{n=0}^{N-2} \left( l_0 + L_n \right)^2 \, .
    \end{aligned}
\end{equation}
Here, $\phi_n$ is a single-component fermionic field obeying the anti-commutation relation $\left\{ \phi^{\dagger}_n, \phi_{n'} \right\}=\delta_{nn'}$. The quantized dimensionless electric field operator $L_n$ and the operator $U_n$ act on the links that connect sites $n$ and $(n+1)$. These two operators obey the commutation relation $\left[U_n, L_{n'}\right]=\delta_{nn'}U_n$. The parameters $m_{\text{lat}}$ and $g$ are the lattice mass and coupling, respectively. The background electric field is given by $l_0 = \theta/\left(2\pi\right)$.

On the lattice, the continuum Gauss's law (\ref{eqn:Gauss_cont}) becomes
\begin{equation} \label{eqn:Gauss_lat}
    L_n - L_{n-1} = Q_n \, ,
\end{equation}
with the charge operator
\begin{equation} \label{eqn:Q_charge_op}
    Q_n = \phi^{\dagger}_n \phi_n - \dfrac{1}{2}(1-(-1)^n) \, .
\end{equation}
Note that the total charge $\sum_{n=0}^{N-1} Q_n$ is conserved due to the global $U(1)$ symmetry of the theory.

Following~\cite{angelides2024firstorder}, we set the electric field on the left boundary to zero. This can be done since a non-zero value can be viewed as a constant additive shift in the value of the background electric field. Using (\ref{eqn:Gauss_lat}), the electric field on each link can then be expressed solely in terms of the fermionic charges as~\cite{angelides2024firstorder}
\begin{equation}
    L_n = \sum\limits_{k=0}^n Q_k \, .
\end{equation}
We can therefore use this to eliminate the gauge fields and the corresponding operators $U_n$~\cite{PhysRevD.56.55}, such that the Hamiltonian in (\ref{eqn:Ham_stagg}) becomes
\begin{equation} \label{eqn:Ham_stagg_int_out}
    \begin{aligned}
        H_{\text{lat}} &= -\dfrac{i}{2a} \sum\limits_{n=0}^{N-2} (\phi^{\dagger}_n \phi_{n+1} - \text{H.c.}) \\
        &+ m_{\text{lat}} \sum\limits_{n=0}^{N-1} (-1)^n \phi^{\dagger}_n \phi_n \\
        &+ \dfrac{a g^2}{2} \sum\limits_{n=0}^{N-2} \left( l_0 + \sum\limits_{k=0}^n Q_k \right)^2 \, .
    \end{aligned}
\end{equation}

To translate the Hamiltonian (\ref{eqn:Ham_stagg_int_out}) to qubits we apply the Jordan-Wigner transformation~\cite{Jordan1928} with
\begin{equation}
    \phi_n = \prod\limits_{k<n} (iZ_k) \dfrac{1}{2} (X_n - i Y_n) \, ,
\end{equation}
where $X_n$, $Y_n$, $Z_n$ are the usual Pauli operators acting on spin $n$. Applying the Jordan-Wigner transformation to the charge operator in~(\ref{eqn:Q_charge_op}) we obtain
\begin{equation}
    Q_n = \dfrac{1}{2}(Z_n + (-1)^n) \, .
\end{equation}
We re-scale the Hamiltonian to make it dimensionless,
\begin{equation}
H_{\text{lat}} \to \dfrac{2}{ag^2} H_{\text{lat}} \, .
\end{equation}
Last, to enforce the condition of vanishing total charge we add a penalty term to the Hamiltonian
\begin{equation} \label{eqn:zero_Qtot_cond}
H_{\text{lat}} \to H_{\text{lat}} + \lambda \left( \sum\limits_{n=0}^{N-1} Q_n \right)^2 \, ,
\end{equation}
while choosing the Lagrange multiplier $\lambda$ sufficiently large.

The final lattice Hamiltonian is then given by~\cite{angelides2024firstorder}
\begin{equation} \label{eqn:Ham_final}
    \begin{aligned}
        H_{\text{lat}} &= \dfrac{x}{2} \sum\limits_{n=0}^{N-2} \left( X_n X_{n+1} + Y_n Y_{n+1} \right) \\
        &+ \dfrac{1}{2} \sum\limits_{n=0}^{N-2} \sum\limits_{k=n+1}^{N-1} \left( N-k-1+\lambda \right) Z_n Z_k \\
        &+ \sum\limits_{n=0}^{N-2} \left( \dfrac{N}{4} - \dfrac{1}{2} \Bigl\lceil \dfrac{n}{2} \Bigr\rceil + l_0 \left( N-n-1 \right) \right) Z_n \\
        &+ \dfrac{m_{\text{lat}}}{g} \sqrt{x} \sum\limits_{n=0}^{N-1} (-1)^n Z_n \\
        &+ l_0^2 \left( N-1 \right) + \dfrac{1}{2} l_0 N + \dfrac{1}{8} N^2 + \dfrac{\lambda}{4} N \, ,
    \end{aligned}
\end{equation}
where $x = 1/\left( ag \right)^2$ is the inverse lattice spacing squared in units of the coupling and $\lceil \cdot \rceil$ is the ceiling function.

The lattice discretization via staggered fermions requires an additive mass renormalization~\cite{PhysRevResearch.4.043133}, such that the renormalized mass is given by
\begin{equation} \label{eqn:mass_renorm}
m_{\text{r}}/g = m_{\text{lat}}/g + \text{MS} \left( V, ag, l_0 \right) \, ,
\end{equation}
where $\text{MS}$ is the additive mass shift and $V = N/\sqrt{x}$ is the dimensionless lattice volume.

For periodic boundary conditions (PBC) the mass shift was analytically computed in~\cite{PhysRevResearch.4.043133} to be
\begin{equation} \label{eqn:MSp}
\text{MS}_{\text{p}} = \dfrac{1}{8\sqrt{x}} \, .
\end{equation}
For OBC the mass shift is not analytically known. However, Ref.~\cite{PhysRevD.108.014516} introduced a method to compute it numerically. In addition, Ref.~\cite{PhysRevD.108.014516} argued that, for sufficiently large volumes $V$, (\ref{eqn:MSp}) provides a good approximation also for OBC. As the mass shift does not play the central role in our analysis, we adopt the latter strategy and use (\ref{eqn:MSp}) for the mass shift in (\ref{eqn:mass_renorm}) while setting $V=30$ for the numerical simulations.

To detect a first-order PT or the absence thereof, we measure the energies of at least two of the lowest eigenstates over the $l_0$ region of interest and observe whether the corresponding levels cross or not when varying $l_0$. For the completeness of our analysis, we also consider two additional observables~\cite{angelides2024firstorder}. One of these is the particle number operator
\begin{equation} \label{eqn:Pnum}
P = \dfrac{N}{2} + \dfrac{1}{2} \sum\limits_{n=0}^{N-1} (-1)^n Z_n \, .
\end{equation}
The other is the electric field density operator in the bulk
\begin{equation} \label{eqn:Lelf}
L = l_0 + \dfrac{1}{4} + \dfrac{1}{2} \sum\limits_{n=0}^{N/2-2} Z_n + \dfrac{1}{4} Z_{N/2-1} \, .
\end{equation}
In the latter, to reduce boundary effects, two neighboring links in the center of the system are considered. Furthermore, the discretization via staggered fermions causes a non-uniform electric flux on the links, due to the staggering of the charge on the lattice. Consequently, to reduce the boundary and the staggering effects, the electric field on two neighboring links in the center of the chain is averaged over~\cite{angelides2024firstorder}.

The commutator of the particle number operator (\ref{eqn:Pnum}) and the Hamiltonian (\ref{eqn:Ham_final}) vanishes in the limit $m_{\text{lat}}/g \gg 1$. Therefore, integer eigenvalues of $P$ represent good quantum numbers for large values of $m_{\text{lat}}/g$~\cite{PhysRevD.96.114501, angelides2024firstorder}. For $m_{\text{lat}}/g \gg 1$ and $\theta < \pi$, the ground state is given by $\ket{10}^{\otimes{N/2}}$, with no particles present. Conversely, for $m_{\text{lat}}/g \gg 1$ and $\theta > \pi$ the state containing a pair of particles, $\ket{11} \otimes \ket{10}^{\otimes{(N-4)/2}} \otimes \ket{00}$, corresponds to the ground state. Here, a negatively (positively) charged particle is localized on the left (right) of the one-dimensional chain at site $1$ ($N-2$)~\cite{angelides2024firstorder}. Panels (b) - (g) of Fig.~\ref{fig:phase_str} show energy $E$, particle number $P$ and electric field density $L$ over $l_0$ for a set of lowest eigenstates for $m_{\text{lat}}/g$ above (panels (b) - (d)) and below (panels (e) - (g)) the second-order PT.

\section{Adiabatic Preparation of States} \label{sect:adiab_st_prep}

In this section we review several versions of the adiabatic theorem and subsequently present our method. We begin by introducing the corresponding necessary conditions that we assume to be valid throughout. First, note that the spectrum of a Hamiltonian on a finite lattice is always discrete. This allows us to apply the versions of the adiabatic theorem stated below.

Second, we assume that the Hamiltonian of a physical system depends sufficiently smoothly on the parameter that is varied during the adiabatic evolution. That is, we assume that the energy levels as functions of the varied parameter are within the $C^{\infty}$ class. Furthermore, we assume that the eigenstates of the system can be degenerate either only at a finite number of parameter values, or on the full range of the considered parameter.

In particular, if eigenstates of an eigensubspace exhibit degeneracy for all values of the parameter varied, we assume the adiabatic theorem to hold analogously for the (absence of) transitions between the different degenerate eigensubspaces, as it does for the (absence of) transitions between the individual eigenstates if no such degeneracy is present.

\subsection{Adiabatic theorem with a gap condition}

The earliest versions of the adiabatic theorem with the condition of a non-vanishing gap were presented in~\cite{ehrenfest1916adiabatische, Born1927, Born1928, 1571980074879151104, Zener1932, Stueckelberg1932, Majorana1932, messiah1961quantum, griffiths2017introduction}. Specifically, consider a quantum system parameterized by a time-dependent Hamiltonian $H(t)$ and prepared in one of its instantaneous eigenstates. The corresponding adiabatic theorem states that the system remains in that state if: 1.\ The Hamiltonian is varied sufficiently slowly, and 2.\ There is a non-zero gap between the eigenvalue of that state and the rest of the spectrum for all $t$.

In particular, for an adiabatic evolution time $T$, the probability to transition into a different state is~\cite{Born1928}
\begin{equation} \label{eqn:adiab_thm_w_gap_p_tr}
    p_{\text{tr}} = O \left[ \left( e T \right)^{-2} \right] \, ,
\end{equation}
where $e$ is a unit of energy.

To provide an intuitive introduction, below we expand upon and quantify the conditions 1.\ and 2.\ to three different degrees of detail.

\subsubsection{First formulation}

At the first-approximation level, there are two characteristic time-scales at play. The first is the typical time-scale $T_{\text{dyn}}$ of the system's intrinsic dynamics determined by the Hamiltonian. The second is the time-scale $T$ on which the Hamiltonian itself is changed externally. Now, the time-scale of the dynamics within the system is governed by and is inversely proportional to the difference in the relevant energy levels. For the remainder of the paper, we define $\Delta$ to be the minimal energy gap between two of the relevant levels.

Using this, we can define a non-adiabaticity parameter,
\begin{equation} \label{eqn:adiab_1st_approx}
    \varepsilon \sim \dfrac{T_{\text{dyn}}}{T} \sim \dfrac{1}{T \Delta} \, ,
\end{equation}
which measures how non-adiabatic the evolution of the state is under a time-dependent Hamiltonian. For the evolution to be adiabatic, $\varepsilon \ll 1$ needs to hold. In other words, the variation in the Hamiltonian is then sufficiently slow, allowing the state to adjust to the changed system. For a given minimal energy gap $\Delta$, we can equivalently express this as a condition on the time-scale $T$ on which the Hamiltonian is to be evolved, $T \gg 1/\Delta$.

\subsubsection{Second formulation}

Having considered the condition of adiabaticity from the perspective of time-scales, we can also study it from the viewpoint of energies. Namely, for the evolution of the state to be adiabatic, the change in the system's energy over the typical time-scale of the intrinsic dynamics $T_{\text{dyn}}$ needs to be much smaller than the minimal energy gap between the relevant levels. That is,
\begin{equation} \label{eqn:adiab_2nd_prelim}
    T_{\text{dyn}} \lvert \braket{\dot{H}} \rvert \ll \Delta \, ,
\end{equation}
where $\braket{\dot{H}} = \bra{\psi(t)} \dot{H} \ket{\psi(t)}$ and $\ket{\psi(t)}$ is the time-evolved state. Note that this version of the adiabaticity condition is more expressive than (\ref{eqn:adiab_1st_approx}), as it includes the information on how the energy of the system in a specific state changes when the Hamiltonian is varied.

Employing the fact that $T_{\text{dyn}} \sim 1/\Delta$ and approximating $\lvert \braket{\dot{H}} \rvert \sim \Delta / T$, from (\ref{eqn:adiab_2nd_prelim}) we can estimate $T \sim \Delta / \lvert \braket{\dot{H}} \rvert \gg 1/\Delta$, recovering the result of the first formulation. That is, we can now express the non-adiabaticity parameter at a higher degree of accuracy as
\begin{equation} \label{eqn:adiab_2nd_approx}
    \varepsilon \sim \dfrac{\lvert \braket{\dot{H}} \rvert}{\Delta^2} \, .
\end{equation}

\subsubsection{Third formulation} \label{subsubsect:adiab_st_prep_gap_3}

We also review a further formulation, variations of which are common in the literature~\cite{farhi2000quantum, PhysRevA.65.042308, 10.1063/1.2798382, Wiebe_2012}. We again denote the Hamiltonian of a quantum system as $H(t)$ and its state as $\ket{\psi(t)}$. We consider the system's evolution according to the Schr\"{o}dinger equation
\begin{equation} \label{eqn:Schr_eqn}
    i \dfrac{\mathrm{d}}{\mathrm{d}t} \ket{\psi(t)} = H(t) \ket{\psi(t)} \, ,
\end{equation}
on a finite time-interval, with $t \in [0,T]$. Here, a parameter of the Hamiltonian is externally varied in time. For example, in application to the Schwinger model lattice Hamiltonian (\ref{eqn:Ham_final}), the parameter that is varied in time is the background electric field, such that $l_0=l_0(t)$.

Now, let $\ket{\tilde{E}_i; t}$ be the instantaneous eigenstates of $H(t)$ with corresponding eigenvalues $\tilde{E}_i(t)$, satisfying
\begin{equation}
    H(t) \ket{\tilde{E}_i; t} = \tilde{E}_i(t) \ket{\tilde{E}_i; t} \, .
\end{equation}
Here, $i$ labels the different eigenstates, with $i = 0$ corresponding to the ground state at all times $t$. Consider, for simplicity, that we prepare the system at $t=0$ in the ground state $\ket{\tilde{E}_0; 0}$. Furthermore, assume that there is a non-zero gap between $\tilde{E}_0(t)$ and the rest of the spectrum for all values of $t$. Then the probability to find the system not in the ground state at time $t \in [0,T]$, i.e.\ the transition probability, is
\begin{equation}
    p_{\text{tr}}(t) = 1 - \lvert \braket{\tilde{E}_0; t | \psi(t)} \rvert^2 \, .
\end{equation}

An estimate of an upper bound on $p_{\text{tr}}(t)$ is commonly given by~\cite{farhi2000quantum, PhysRevA.65.042308}
\begin{equation}
     p_{\text{tr}}(t) \lesssim \left( \max_{\substack{t \in [0,T] \\ i \neq 0}} \dfrac{\lvert \bra{\tilde{E}_i; t} \frac{\mathrm{d}H}{\mathrm{d}t} \ket{\tilde{E}_0; t} \rvert}{\left(\tilde{E}_i(t)-\tilde{E}_0(t)\right)^2} \right)^2 \, .
\end{equation}

To make a connection to the second formulation in (\ref{eqn:adiab_2nd_approx}), we can estimate
\begin{equation} \label{eqn:adiab_thm_w_gap_p_tr_bound}
    p_{\text{tr}}(t) \lesssim \left( \dfrac{\lvert \braket{\dot{H}} \rvert}{\Delta^2} \right)^2 \sim \varepsilon ^2 \, .
\end{equation}
That is, in this formulation, from $\varepsilon \ll 1$ we indeed recover the simplified adiabaticity condition $T \gg 1/\Delta$ as well as a suppressed transition probability $p_{\text{tr}}(t) \ll 1$. We also note that the bound on the transition probability in (\ref{eqn:adiab_thm_w_gap_p_tr_bound}) is consistent with (\ref{eqn:adiab_thm_w_gap_p_tr}).

So far we have considered systems with a gap condition. However, can a statement be made for systems where the gap is zero?

\subsection{Adiabatic theorem with level crossings}

\subsubsection{The theorem}

Consider now a system where a degeneracy of the energy levels is caused by crossing. An adiabatic theorem that applies to such systems was developed in~\cite{Born1928}. We first remark that, in contrast to the index $i$ of the instantaneous eigenstate $\ket{\tilde{E}_i;t}$ (see section~\ref{subsubsect:adiab_st_prep_gap_3}), the index $\alpha$ labels the energy level $E_{\alpha}$ corresponding to $\ket{\alpha}$, and that each such level preserves its index throughout, irrespective of whether it crosses a different energy level $E_{\beta}$, $\alpha \neq \beta$, or not. We also introduce a dimensionless parameter
\begin{equation}
    \tau = \dfrac{t}{T} \, .
\end{equation}

By the argument of~\cite{Born1928}, if\footnote{In addition to mild supplementary conditions on boundedness and monotonicity of other quantities, for details see~\cite{Born1928}.} each function $(E_{\alpha}(\tau)-E_{\beta}(\tau))$, $\alpha \neq \beta$, has a finite number of zeros of maximally the $r$-th order and in the vicinity of each zero point $\tau_0$ the estimation
\begin{equation}
    \dfrac{e}{\lvert E_{\alpha}(\tau)-E_{\beta}(\tau) \rvert} < \dfrac{C}{\lvert \tau-\tau_0 \rvert^{r}}
\end{equation}
holds, where $C$ is a dimensionless constant (see~\cite{Born1928} for details), then the transition probability to a different energy level satisfies the expression
\begin{equation} \label{eqn:adiab_thm_w_cr_p_tr}
    p_{\text{tr}} = O \left[ \left( e T \right)^{-2/(r+1)} \right] \, .
\end{equation}
Note that if $r=0$, i.e.\ the levels do not cross, we obtain the transition probability in (\ref{eqn:adiab_thm_w_gap_p_tr}). Thus, the gap condition is not necessary for the adiabatic theorem to hold, and the time-dependence of the transition probability is instead controlled by the slope of the levels. Many works have since expanded on the adiabatic theorem, see e.g.~\cite{doi:10.1143/JPSJ.5.435, HAGEDORN1989278, Avron1999, Teufel2001, PhysRevA.71.063409, Avron2012}.

\subsubsection{Application to the Schwinger model}

We can thus apply the adiabatic theorem of~\cite{Born1928} to the lattice Schwinger model in (\ref{eqn:Ham_final}). In particular, we can employ it to study both the sub-critical mass regime where there is a gap, and the super-critical mass regime where there is a level crossing.

Specifically, consider a range of $l_0$ given by $[l_0^{\text{min}}, l_0^{\text{max}}]$. Take the state $\ket{\alpha}$ corresponding to a level $E_{\alpha}$ at $l_0^{\text{min}}$ as the initial state at $t=0$. Then evolve this state over $t \in [0, T]$ according to the Schr\"{o}dinger equation (\ref{eqn:Schr_eqn}) with the Hamiltonian $H(t)=H_{\text{lat}}(l_0(t))$ from (\ref{eqn:Ham_final}), where $l_0=l_0(t)$ is now time-dependent and is parametrized by the linear ramp
\begin{equation} \label{eqn:l0_lin_ramp}
    l_0(t) = l_0^{\text{min}} + \dfrac{l_0^{\text{max}} - l_0^{\text{min}}}{T} t \, .
\end{equation}
Then, by the adiabatic theorem of~\cite{Born1928}, provided the evolution time $T$ is sufficiently large, the probability to transition to a different level $E_{\beta}$, $\alpha \neq \beta$ is negligible, even when passing through a level crossing.

\subsection{Proposed methodology} \label{subsect:prop_method}

To summarize, consider a system on a finite lattice for which we possess no knowledge about its phase diagram (see section~\ref{sect:disc} for a more detailed discussion). Suppose we would like to study the system's phase structure over the range $[p_{\text{min}}, p_{\text{max}}]$ of a parameter $p$ on which the Hamiltonian of the model $H(p)$ depends. We let the parameter $p$ depend on time $t \in [0, T]$, so that $p(t=0)=p_{\text{min}}$ and $p(t=T)=p_{\text{max}}$. We individually evolve the lowest energy levels with a time-dependent Hamiltonian $H(p(t))$ over the parameter region of interest. By the adiabatic theorem in~\cite{Born1928}, provided $T$ is sufficiently large, we can then recover the phase structure of the model.

That is, we would detect a PT if the two lowest energy levels cross, or the absence thereof if they do not. If the levels cross, we can infer the order of the PT as follows: We first fit each of the levels with a polynomial in the vicinity of the crossing point. We then compute the derivatives of the individual levels with respect to the parameter $p$. A discontinuity in the $n$-th derivative of the energy of the ground state $\tilde{E}_0(p)$ at the crossing point, while all $k$-th derivatives of $\tilde{E}_0(p)$ for $0 \leq k < n$ are continuous, would indicate a PT of $n$-th order.

In general, a system could have an intricate level structure in the parameter range of interest. For example, more than two energy levels could assume the role of the ground state over different values of the parameter $p$. Recovering the corresponding phase structure would require evolving more than two levels. Nevertheless, the approach remains valid and the phase diagram of any finite lattice system can thus be recovered. Below we formalize and quantify our method.

\subsection{Adiabatic evolution of multiple states} \label{subsect:adiab_ev_mult_st}

Here we present our method, as well as a specific version of it implemented in Algorithm~\ref{alg:adiab_ev_mult_st}. We first introduce the algorithm and subsequently expand on the general approach. Throughout Algorithm~\ref{alg:adiab_ev_mult_st}, when it is executed on a quantum device or simulator, we assume that the time-step size, the Suzuki-Trotter order of the decomposition as well as the number of shots are chosen such that the evolution depends sufficiently weakly on the variation of these parameters.

\subsubsection{The algorithm}

The input of the algorithm consists of: First, the parameter range $[p_{\text{min}}, p_{\text{max}}]$ that is to be studied. Second, the time-dependent Hamiltonian of the system $H(p(t))$, which includes the dependence of the parameter $p$ on time, according to the chosen ramp. The ramp sweeps over the parameter range, such that $p(t=0)=p_{\text{min}}$ and $p(t=T)=p_{\text{max}}$. Third, a set of lowest-energy eigenstates $\{ \ket{A} \}$ at the boundary of the parameter range $p_{\text{min}}$. Fourth, an initial value for the total evolution time $T_{\text{in}}$. Fifth, a strictly monotonically increasing function $f:T \mapsto T'$, $T' > T$, which determines how the total evolution time is increased for each iteration of the algorithm. Sixth, a user-defined bound $\mathcal{B}$ on the normalized mean Euclidean point-wise distance over $p$ of the same energy level between two consecutive iterations of the algorithm, summed over all considered energy levels.

We define this norm between two computations of the same energy levels $E_{\alpha}(p)$, corresponding to two consecutive iterations of the algorithm, as
\begin{equation} \label{eqn:alg_Eucl_norm}
    \mathcal{E} = \dfrac{1}{\epsilon \sqrt{\mathcal{N}}} \sum_{\alpha} \lVert E_{\alpha}^{\text{old}}(p) - E_{\alpha}(p) \rVert \, .
\end{equation}
Here $\alpha$ labels the energy level corresponding to $\ket{\alpha} \in \{ \ket{A} \}$. The Euclidean norm is given by
\begin{equation}
    \lVert E_{\alpha}^{\text{old}}(p) - E_{\alpha}(p) \rVert = \sqrt{\sum\limits_{i=1}^{\mathcal{N}} \left( E_{\alpha}^{\text{old}}(p_i) - E_{\alpha}(p_i) \right)^2} \, .
\end{equation}

To calculate this norm, the result $E_{\alpha}(p)$ of the new iteration is interpolated over $p$, and evaluated over the set of $\mathcal{N}$ parameter points of the old result $E_{\alpha}^{\text{old}}(p)$. Note that $\mathcal{N}$ depends on the iteration of the algorithm with
\begin{equation}
    \mathcal{N} = \dim (\{ p \}) ~ \text{of} ~ E_{\alpha}^{\text{old}}(p) \, .
\end{equation}
In (\ref{eqn:alg_Eucl_norm}) we normalize with respect to the difference
\begin{equation} \label{eqn:Eucl_norm_1st_iter}
    \epsilon = \dfrac{1}{\sqrt{\mathcal{N}}} \sum_{\alpha} \lVert E_{\alpha}^{1\text{st}}(p) - E_{\alpha}^{2\text{nd}}(p) \rVert
\end{equation}
of the respective levels between the first and the second iterations of the algorithm. Here $\mathcal{N} = \dim (\{ p \})$ of $E_{\alpha}^{1\text{st}}(p)$.

The algorithm terminates when the Euclidean norm becomes smaller than the user-set total error bound, $\mathcal{E} < \mathcal{B}$. Note that this does not guarantee that the levels converge to their true dependence on $p$. Instead, this means that the dependence of the levels on $p$ is not affected by the current value of $T$ sufficiently enough for $\mathcal{E} \geq \mathcal{B}$ to hold.

Within the algorithm, in order to compute the value and the position of the minimal energy gap, we define the level $E_{\gamma}(p)$ as that corresponding to the ground state at $p_{\text{min}}$. Finally, the output of the algorithm is as follows: First, the final total evolution time $T$, for which $\mathcal{E} < \mathcal{B}$ is satisfied. Second, the value of the minimal energy gap $\Delta$. Third, the corresponding parameter value $p_{\Delta}$. That is, $p_{\Delta}$ is such that the gap has a value of $\Delta$ at $p_{\Delta}$.

\begin{algorithm}[htbp]
\SetAlgoLined
\KwIn{Parameter range $[p_{\text{min}}, p_{\text{max}}] \ni p$, Hamiltonian $H(p(t))$, set of lowest-energy eigenstates $\{ \ket{A} \}$ at $p_{\text{min}}$, initial value for total evolution time $T_{\text{in}}$, strictly monotonically increasing function $f:T \mapsto T'$ with $T' > T$, error bound $\mathcal{B}$}
\KwOut{Final evolution time $T$, minimal gap $\Delta$ and corresponding parameter value $p_{\Delta}$}
\BlankLine
$T \leftarrow T_{\text{in}}$\;
$\mathcal{E} \leftarrow \mathcal{B}$\;
\While{$\mathcal{E} \geq \mathcal{B}$}{
    \For{$\ket{\alpha} \in \{ \ket{A} \}$}{
        evolve $\ket{\alpha}$ with $H(p(t))$ for $t \in [0, T]$\;
        $E_{\alpha}(p) \leftarrow \bra{\alpha(t)} H(p(t)) \ket{\alpha(t)}$\;
        }
    \If{$T > T_{\text{in}}$}{
        \If{$T = f(T_{\text{in}})$}{
            $\epsilon \leftarrow \dfrac{1}{\sqrt{\mathcal{N}}} \sum\limits_{\alpha} \lVert E_{\alpha}^{\text{old}}(p) - E_{\alpha}(p) \rVert$\;
            }
        $\mathcal{E} \leftarrow \dfrac{1}{\epsilon \sqrt{\mathcal{N}}} \sum\limits_{\alpha} \lVert E_{\alpha}^{\text{old}}(p) - E_{\alpha}(p) \rVert$\;
        }
    $T \leftarrow f(T)$\;
    $E_{\alpha}^{\text{old}}(p) \leftarrow E_{\alpha}(p)$\;
    $\mathcal{N} \leftarrow \dim (\{ p \})$ of $E_{\alpha}^{\text{old}}(p)$\;
    }
$\gamma \leftarrow \alpha'$ such that $E_{\alpha'}(p_{\text{min}}) = \min\limits_{\ket{\alpha} \in \{ \ket{A} \}}\left[ E_{\alpha}(p_{\text{min}}) \right]$\;
\eIf{$E_{\gamma}(p)$ crosses with any other $E_{\alpha}(p)$ for $\alpha \neq \gamma$}{
    $\Delta \leftarrow 0$\;
    find $p_{\Delta}$ by interpolating corresponding levels\;
    }{
    $\Delta \leftarrow \min\limits_{\substack{p \in [p_{\text{min}}, p_{\text{max}}] \\ \ket{\alpha} \in \{\ket{A}\} ,~\ket{\alpha} \neq \ket{\gamma}}} \lvert E_{\alpha}(p)-E_{\gamma}(p) \rvert$\;
    $p_{\Delta} \leftarrow \operatorname*{arg\,min}\limits_{\substack{p \in [p_{\text{min}}, p_{\text{max}}] \\ \ket{\alpha} \in \{\ket{A}\},~\ket{\alpha} \neq \ket{\gamma}}} \lvert E_{\alpha}(p)-E_{\gamma}(p) \rvert$\;
    }
\caption{Evolution of multiple states}
\label{alg:adiab_ev_mult_st}
\end{algorithm}

When evaluating an iteration of the algorithm for any finite value of $T$, irrespective of how large, it is impossible to claim that this value of $T$ is sufficiently large to guarantee that the dependence of the energy levels on $T$ has reached its asymptotic behavior. Therefore, applying the algorithm to a system where we do not know the phase diagram, we make the following key assumption: If we observe convergence of the individual energy levels from iteration to iteration of the algorithm (i.e.\ increasing $T$), we assume that this convergence is asymptotic and the individual energy levels approach their true dependence on the parameter $p$.

To summarize, we assume
\begin{asmp} \label{asmp:asympt_assumpt}
$\mathcal{E} \to 0 \quad \Rightarrow \quad T$ sufficiently large.
\end{asmp}

\subsubsection{The general approach}

At its core, our method is a direct application of the adiabatic theorem. Assume that we are given a Hamiltonian and the parameter range of the ``interesting'' regime, where one of the following occurs for the two lowest levels: A single level crossing, a single common point of the levels without a level crossing, or a minimal non-zero gap. However, we are not given the information about which of the above it is and for which value of the parameter it occurs. Assume also that all higher levels possess much higher energies than the lowest two on the entire parameter range. We proceed as follows: We find the two lowest eigenstates on one boundary of the parameter range and evolve the states separately with a time-dependent Hamiltonian for a (relatively short) chosen time $T$, sweeping over the parameter range.

We then increase the evolution time $T$ and repeat the procedure. We continue iterating the scheme over greater values of $T$ until the difference between two consecutive computations of the levels $\mathcal{E}$ becomes smaller than the user-set bound $\mathcal{B}$. If we observe that $\mathcal{E}$ approaches zero, we assume that the evolution time $T$ is sufficiently large (otherwise we set a lower bound $\mathcal{B}$ and re-run the algorithm). Then, under this assumption, by the adiabatic theorem of~\cite{Born1928}, we have found how the energy levels of the model depend on the chosen parameter. From this parameter-dependence of the energy levels, we can then extract the information about the phase diagram of the model as described in section~\ref{subsect:prop_method}.

Moreover, under the assumption~\ref{asmp:asympt_assumpt}, if we obtain a result for a certain parameter choice (e.g.\ a specific value of $m_{\text{lat}}/g$ for the Schwinger model), we can estimate a value of $T$ for a neighboring parameter choice. Note that iterating the algorithm on a quantum device, for each value of $T$ it is necessary to vary the time-step size $\delta t$, the Suzuki-Trotter order of the decomposition, and the number of shots $n_{\text{shots}}$, until the change due to this variation is sufficiently small (see also the comment in section~\ref{subsubsect:num_an_add_rem}).

In case we are not given any information except for the Hamiltonian, we may need to evolve more than two levels very slowly and over large parameter ranges. However, on the contrary, if we are provided with additional information, our method correspondingly gains more predictive power. We discuss this in more detail in section~\ref{sect:disc}.

In particular, if we know the true phase structure of the model (e.g.\ by exact diagonalization), we can test our method (see section~\ref{sect:num_an} for numerical results). To do this, we adapt the algorithm: Instead of (\ref{eqn:alg_Eucl_norm}), we calculate the norm with respect to the exact diagonalization (ED) results for each individual iteration,
\begin{equation} \label{eqn:alg_ED_Eucl_norm}
    \tilde{\mathcal{E}} = \dfrac{1}{\tilde{\epsilon} \sqrt{\mathcal{N}}} \sum_{\alpha} \lVert E_{\alpha}^{\text{ED}}(p) - E_{\alpha}(p) \rVert \, .
\end{equation}
Here, $\epsilon$ from (\ref{eqn:Eucl_norm_1st_iter}) is accordingly modified to
\begin{equation}
    \tilde{\epsilon} = \dfrac{1}{\sqrt{\mathcal{N}}} \sum_{\alpha} \lVert E_{\alpha}^{\text{ED}}(p) - E_{\alpha}^{1\text{st}}(p) \rVert \, ,
\end{equation}
and $\mathcal{N}$ is now given by $\mathcal{N} = \dim (\{ p \})$ of $E_{\alpha}(p)$, with the Euclidean norm evaluated over the set of points of $E_{\alpha}(p)$, correspondingly.

\subsubsection{The preparation of lowest-energy eigenstates}

The final aspect of the method to address is the preparation of the set of lowest-energy eigenstates $\{ \ket{A} \}$ at $p_{\text{min}}$. This procedure is model-specific. First, as a supplementary technique, to restrict the set of eigenstates to be considered, it is possible to study the low-energy eigenspectrum for small-size systems on a classical device and use this to infer the low-energy eigenspectrum for large systems to be studied on quantum devices. Now, the general approach we use consists of two steps: First, pick a region of the parameter space where the eigenspectrum is known. Second, apply the method above to evolve the eigenstates to the target parameter region.

In the specific example of the lattice Schwinger model in (\ref{eqn:Ham_final}) for an even number of sites $N \geq 4$, for fixed values of $m_{\text{lat}}/g$, $l_0 = l_0^{\text{min}}$ and $\lambda$, we first consider only the terms of the Hamiltonian containing the $Z_n$ and $Z_n Z_k$ operators (as well as the term containing the global identity operator). The eigenspectrum is then given by the $2^N$ states $\ket{s}^{\otimes{N}}$ with $ \ket{s} \in \{ \ket{0}, \ket{1} \}$. We now select the states by their total charge $\sum_{n=0}^{N-1} Q_n$. Consider two subsets of eigenstates $\{ \ket{A} \}$ and $\{ \ket{B} \}$ belonging to sectors of total charges $Q_A^{\text{tot}}, Q_B^{\text{tot}} = -N/2, \dotsc, N/2$, respectively, with $|Q_A^{\text{tot}}| < |Q_B^{\text{tot}}|$. By (\ref{eqn:zero_Qtot_cond}), the eigenstates from subset $\{ \ket{A} \}$ will have energies lower than those of eigenstates from subset $\{ \ket{B} \}$ by approximately $\lambda [(Q_B^{\text{tot}})^2 - (Q_A^{\text{tot}})^2]$.

If $\lambda$ is sufficiently large, the same will hold to a good approximation for a range of values for the physical parameters of the model. In particular, the energy difference between the sectors $Q_A^{\text{tot}} = 0$ and $Q_B^{\text{tot}} = 1$ will be approximately $\lambda$. That is, the eigenstates in our target sector of vanishing total charge will have energies lower by approximately $\lambda$ than those with unit absolute total charge.

We then apply our method and evolve the system such that the remaining parameter $x$, which parametrizes the term containing $X_n X_{n+1}$ and $Y_n Y_{n+1}$ operators, acquires the desired target value and we obtain the full lattice Schwinger model (\ref{eqn:Ham_final}).

In the regime of sufficiently large $m_{\text{lat}}/g$, by arguments in section~\ref{subsect:lat_form}, we can further restrict the set of low-energy eigenstates by considering only the two states that are well approximated by $\ket{10}^{\otimes{N/2}}$ and $\ket{11} \otimes \ket{10}^{\otimes{(N-4)/2}} \otimes \ket{00}$ with $P=0$, $L=l_0$ and $P=2$, $L=l_0 - 1$, correspondingly. In the numerical analysis below, for demonstration purposes, we restrict ourselves to two eigenstates prepared by ED at $l_0^{\text{min}}$. However, as detailed above, the states can be prepared via adiabatic evolution.

\section{Numerical Analysis} \label{sect:num_an}

\subsection{The set-up}

\subsubsection{Specifications of the simulations}

For the numerical analysis of the Schwinger model lattice Hamiltonian in (\ref{eqn:Ham_final}) we perform three types of simulations: First, ED using QuSpin~\cite{10.21468/SciPostPhys.2.1.003, 10.21468/SciPostPhys.7.2.020, QuSpin_doc}. Second, adiabatic state preparation using QuTiP~\cite{JOHANSSON20121760, JOHANSSON20131234, QuTiP_ex}, which we call ``time-evolution on a classical device''. Third, adiabatic state preparation using Qiskit~\cite{qiskit2024} on a classical computer which we call ``time-evolution on a digital quantum simulator''.

The input parameters of the model common for all three types of simulations are the following: The total site number is $N = 6$, the dimensionless lattice volume is $V = 30$ and the Lagrange multiplier is $\lambda = 100$. We found this value of $\lambda$ to be sufficient to ensure a vanishing total charge in our simulations. The ED simulations in Figs.~\ref{fig:phase_str}, \ref{fig:m_high} and \ref{fig:m_low} are performed with $10^3$ steps. For the adiabatic state preparation simulations via (\ref{eqn:Schr_eqn}) we employ the linear ramp from (\ref{eqn:l0_lin_ramp})\footnote{As an additional remark, to further optimize our method, it is possible to modify the linear ramp of (\ref{eqn:l0_lin_ramp}) and instead consider a local ramp, such as that in~\cite{sym14040809}, with an additional lower bound on the ramp speed for the case of level crossings.}. For the strictly monotonically increasing function $f:T \mapsto T'$, $T' > T$, we choose the following sequence: For the first nine iterations $i$ of the algorithm we take
\begin{equation}
     T_i = 5 i \, , ~ 1 \leq i \leq 9 \, . 
\end{equation}
For the subsequent iterations we use the recursion relation
\begin{equation}
    T_i = 10 T_{i-9} \, , ~ i \geq 10 \, .
\end{equation}
The time-step size is constant throughout and is given by $\delta t = 0.5$. This results in the number of time-steps given by $n_{\delta t, i} = 2 T_i$ for each corresponding simulation. That is, we explore each order of magnitude for the number of time-steps linearly. The parameter $l_0$ is varied more slowly for a greater value of $T$. The $l_{0, i}$-step size for an algorithm iteration $i$ is given by $\delta l_{0,i} = (l_0^{\text{max}}-l_0^{\text{min}})/n_{\delta t, i}$.

The time-evolution on a digital quantum simulator is subject to shot-induced noise, with $n_{\text{shots}} = 10^4$ shots per circuit. Furthermore, for the quantum simulator we set the order of the Suzuki-Trotter decomposition~\cite{Hatano2005, Berry2007} equal to $q_{\text{ST}}=6$ throughout.

\subsubsection{Additional remarks} \label{subsubsect:num_an_add_rem}

Note that the time-step size $\delta t$, the $l_0$-step size $\delta l_0$, the number of steps $n=n_{\delta t}=n_{\delta l_0}$, the total evolution time $T=n \, \delta t$ and the Suzuki-Trotter order $q_{\text{ST}}$ are all co-dependent parameters. That is, to make the parameter sweep more adiabatic by increasing $T$ while keeping $\delta t$ fixed, we need to increase $n$. Although this reduces $\delta l_0$ for a fixed range $[l_0^{\text{min}}, l_0^{\text{max}}]$, this also increases the error of the Trotter expansion (TE) after time $T$, with the corresponding error given by $\epsilon_{\text{TE}} = T (\delta t)^{q_{\text{ST}}}$. To prevent this error from increasing, while maintaining the new value for $T$ with a fixed $\delta t$, the Suzuki-Trotter order $q_{\text{ST}}$ needs to be increased.

In general, in addition to the time-evolution parameters, we expect the value of a sufficiently large $T$ to also depend on the model considered, its size in terms of the number of qubits $N$, and the chosen model parameters.

For completeness, in addition to the eigenstate energy $E$, we also compute the particle number $P$ from (\ref{eqn:Pnum}) and the electric field density $L$ from (\ref{eqn:Lelf}) for all three types of simulations. Below we present the results of our numerical analysis.

\subsection{Regime above the second-order phase transition}

\begin{sidewaysfigure*}[htbp]
    \centering
    \vspace{10cm}
    \includegraphics[width=\textwidth]{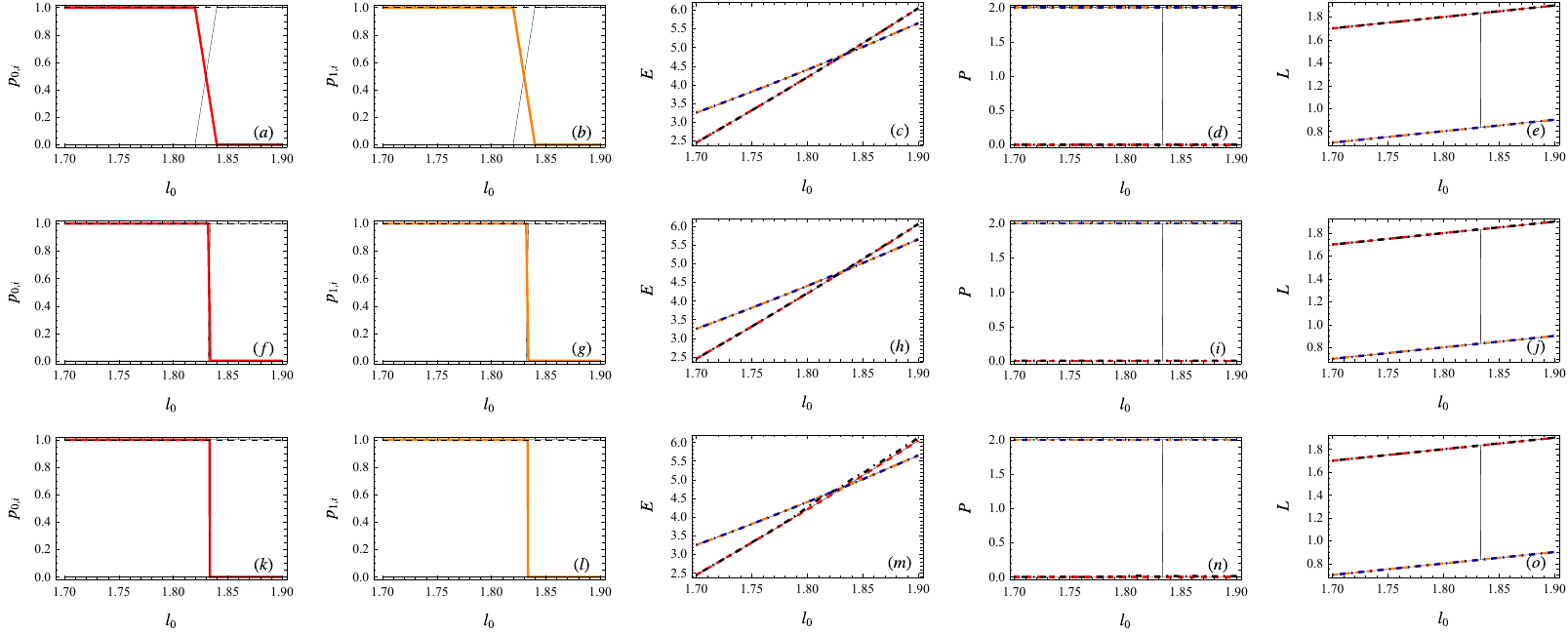}
    \caption{\small First-order phase transition in the regime above the second-order phase transition. Here we set $m_{\text{lat}}/g = 10$. Top, middle and bottom rows show the results for $T=5$, $50$ and $500$, respectively. The first column (from the left) shows the time-evolution of the ground state at $l_0^{\text{min}}=1.7$ evolved according to (\ref{eqn:Schr_eqn}) with (\ref{eqn:Ham_final}) and (\ref{eqn:l0_lin_ramp}) from $l_0^{\text{min}}=1.7$ to $l_0^{\text{max}}=1.9$ on a classical device. The respective panels show the probability $p_{0,i}$, plotted over $l_0$, to find the evolved state in the various instantaneous eigenstates $\ket{\tilde{E}_i;t}$ at a corresponding $l_0$. The probability to remain in the ground state is shown in solid red. The probabilities to transition to each of the three lowest excited states depicted in Figs.~\ref{fig:phase_str}(b) - \ref{fig:phase_str}(d) are shown in thin solid black. The total probability of the above four lowest eigenstates is shown in thin dashed black. The second column shows the analogous probability $p_{1,i}$ for the evolved first excited state at $l_0^{\text{min}}=1.7$. The probability to remain in the first excited state is shown in solid orange. Third, fourth and fifth columns show the eigenstate energy $E$, particle number $P$ and electric field density $L$ over $l_0$, respectively. The results of exact diagonalization for the two lowest states are shown in thin solid black. The results for the evolution of the two lowest states at $l_0^{\text{min}}=1.7$ on a classical device (digital quantum simulator) are shown in dashed red and dashed orange (dash-dotted black and dash-dotted blue), respectively.}
    \label{fig:m_high}
\end{sidewaysfigure*}

The numerical results for $m_{\text{lat}}/g = 10$ for the sub-set of the total evolution times $T \in \{ 5, 50, 500 \}$ are shown in Fig.~\ref{fig:m_high}. We observe that the two lowest levels cross (see section~\ref{subsect:conv_an} for the continuation of this discussion). We find indications that the first derivative of the energy of the ground state $\tilde{E}_0(l_0)$ is discontinuous at the crossing point $l_0^*$. Correspondingly, we indeed recover the first-order PT. This agrees with the ED results. Furthermore, note from the ED results shown in Fig.~\ref{fig:phase_str}(b) that the higher excited states have a much higher energy, making the probability to transition into them negligible. This is verified by the results shown in the first and second columns of Fig.~\ref{fig:m_high}. The former shows the probability $p_{0,i} = \lvert \braket{\tilde{E}_i;t | \psi(t)} \rvert^2$ to find the ground state $\ket{\tilde{E}_0;0}$ at $l_0^{\text{min}}$ evolved under (\ref{eqn:Schr_eqn}) in the instantaneous eigenstate $\ket{\tilde{E}_i;t}$ of $H(t)$ at a time $t$. The latter similarly shows $p_{1,i} = \lvert \braket{\tilde{E}_i;t | \psi(t)} \rvert^2$ for the evolved first excited state $\ket{\tilde{E}_1;0}$ at $l_0^{\text{min}}$.

Note from the bottom row of Fig.~\ref{fig:m_high} that for $T = 500$ the results on a digital quantum simulator (shown in dash-dotted black and dash-dotted blue) deviate slightly from the results on a classical device (shown in dashed red and dashed orange). This indicates that the time-step $\delta t$ needs to be reduced and/or the number of shots $n_{\text{shots}}$ needs to be increased (see the discussion in section~\ref{subsect:conv_an}).

Nevertheless, we find that an evolution time of $T=5$ is already sufficient to recover the dependence of the energy levels on $l_0$, as well as that of $P$ and $L$ of the corresponding levels, as the results in the rows for $T=5$, $50$ and $500$ are almost the same. This is supported by the observed slopes of the energy levels: We find that for the two relevant levels $E_{\alpha}$ and $E_{\beta}$, $\alpha \neq \beta$, the function $(E_{\alpha}(\tau)-E_{\beta}(\tau))$ has a single zero of order $r = 1$. This is the lowest possible order for a crossing. Note from~(\ref{eqn:adiab_thm_w_cr_p_tr}) that the transition probability decreases with smaller $r$, consistent with our results.

We are therefore able to successfully validate the method described in section~\ref{subsect:adiab_ev_mult_st} in a particular setting. Namely, our method recovered the previously known result of a level crossing and a corresponding first-order PT on the given range of $l_0$. Moreover, we were able to recover the corresponding critical value $l_0^*$ where the first-order PT takes place. Below we first discuss the results of our numerical analysis in the no-transition region of the phase diagram. We then consider the dependence of the errors $\mathcal{E}$ and $\tilde{\mathcal{E}}$ on the iteration of the algorithm for both regimes.

\subsection{Regime below the second-order phase transition}

\begin{sidewaysfigure*}[htbp]
    \centering
    \vspace{10cm}
    \includegraphics[width=\textwidth]{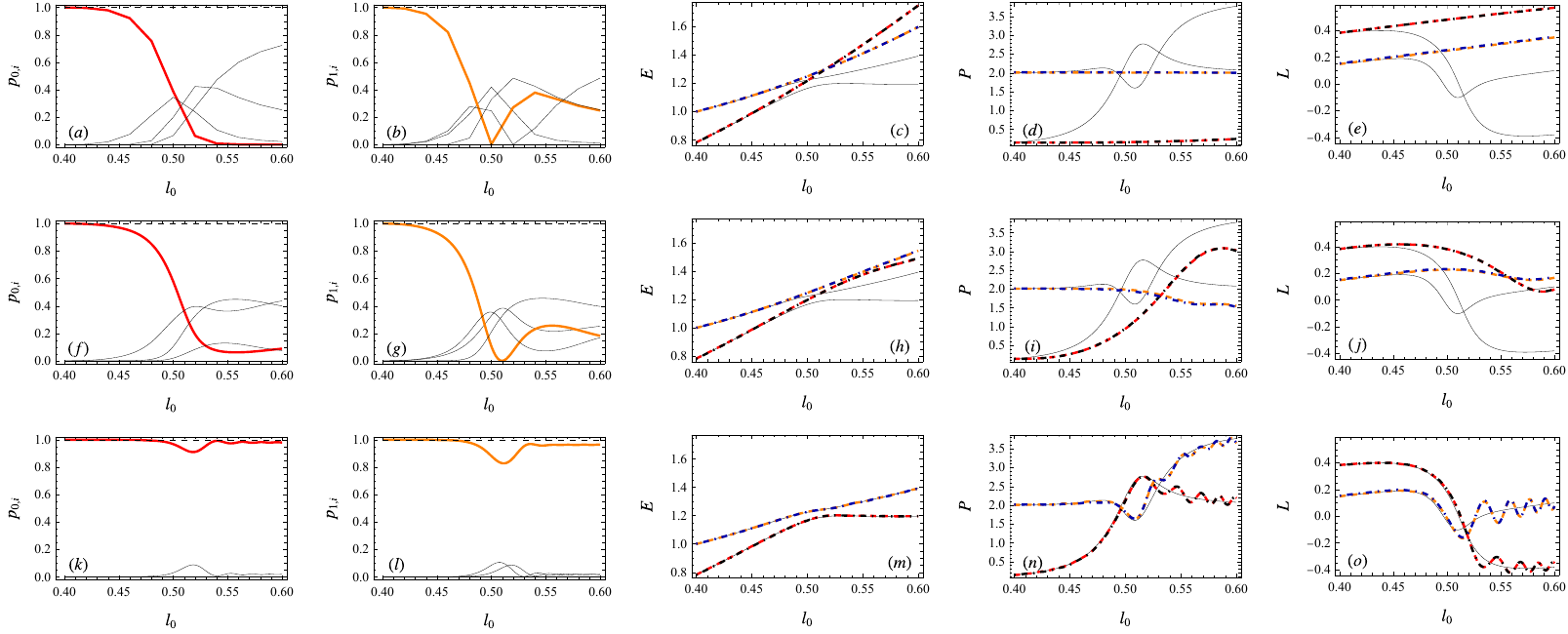}
    \caption{\small No phase transition in the regime below the second-order phase transition. Here we set $m_{\text{lat}}/g = 0$. For the panels in the first and second columns we consider the five lowest eigenstates depicted in Figs.~\ref{fig:phase_str}(e) - \ref{fig:phase_str}(g). We evolve the two lowest eigenstates at $l_0^{\text{min}}=0.4$ according to (\ref{eqn:Schr_eqn}) with (\ref{eqn:Ham_final}) and (\ref{eqn:l0_lin_ramp}) from $l_0^{\text{min}}=0.4$ to $l_0^{\text{max}}=0.6$. In all other aspects the description follows that of Fig.~\ref{fig:m_high}.}
    \label{fig:m_low}
\end{sidewaysfigure*}

The results for the regime $m_{\text{lat}}/g = 0$ are presented in Fig.~\ref{fig:m_low}. We observe that as $T$ is increased, the dependence of the two lowest levels on $l_0$ obtained by our method approaches the ED results. Specifically, we successfully recover the non-zero gap between the two lowest levels. Here, the slow convergence of the adiabatic results towards those of ED is due to the proximity in energy of the higher excited states to the two lowest levels (see Fig.~\ref{fig:phase_str}(e)). In particular, for small values of $T$ the states evolve into superpositions that include higher excited states. Furthermore, for small $T$ we observe a crossing of the levels. However, for sufficiently large $T$ we converge towards the true ED results.

In this example, again, using our adiabatic method, we successfully recover the known dependence of the energy levels on $l_0$. Namely, we verify that indeed there are no PTs on this range of $l_0$. Moreover, we are able to find the minimal value of the non-zero energy gap between the ground state and the rest of the spectrum on this $l_0$-range.

The fact that the time-evolution has been sufficiently adiabatic is also supported by the values we obtain for the probabilities to remain in the corresponding initial states; $p_{0,i=0}(T) \approx 0.98$ and $p_{1,i=1}(T) \approx 0.96$ from Figs.~\ref{fig:m_low}(k) and \ref{fig:m_low}(l), respectively.

Although not perfectly consistent with adiabaticity, the minimal probabilities (on the entire $l_0$ range) to remain in the initial states $p_{0,i=0}^{\text{min}} \approx 0.91$ and $p_{1,i=1}^{\text{min}} \approx 0.83$ are nevertheless sufficiently large to ensure that for each individual simulation, at the end of the time-evolution at $t=T$, the system is found in a state that lies in the proximity of the level corresponding to the initial state. This is because the energy gap increases again as $l_0$ approaches $l_0(T)=l_0^{\text{max}}$. In addition, the oscillations of $P$ and $L$ in Figs.~\ref{fig:m_low}(n) and \ref{fig:m_low}(o), respectively, result from an imperfect overlap of the desired ideal target eigenstate obtained from ED and the actual evolved state (cf.\ Figs.~\ref{fig:m_low}(k) and \ref{fig:m_low}(l)).

\subsection{Convergence analysis} \label{subsect:conv_an}

\begin{figure*}[htbp]
    \centering
    \includegraphics[width=\textwidth]{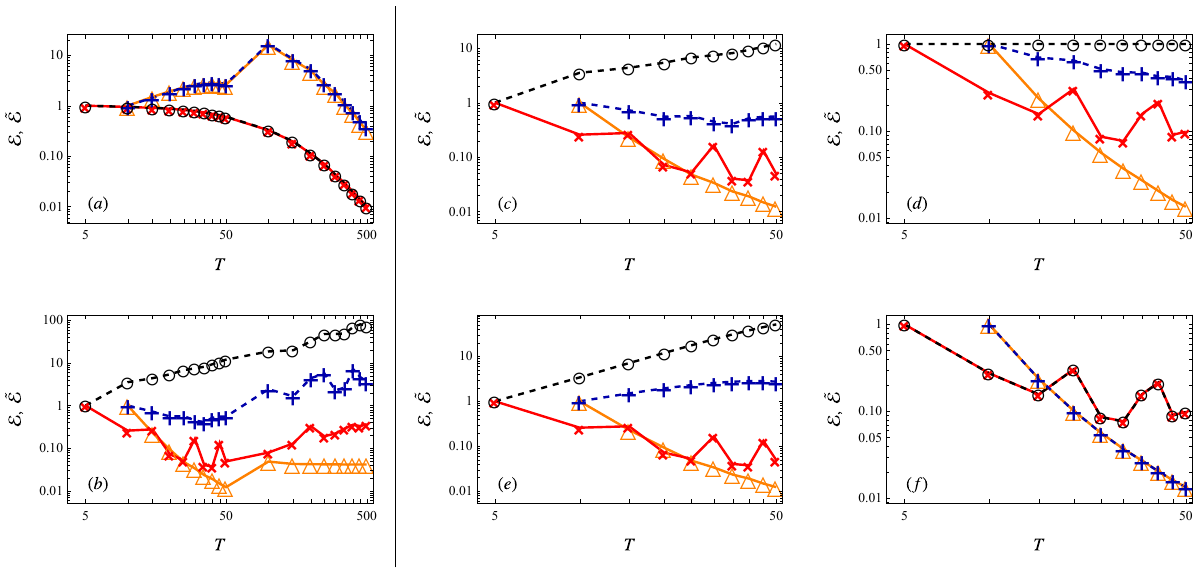}
    \caption{\small Errors $\mathcal{E}$ and $\tilde{\mathcal{E}}$ over evolution time $T$. The errors $\mathcal{E}$, calculated from (\ref{eqn:alg_Eucl_norm}), between two consecutive iterations on a classical device and on a digital quantum simulator are shown as orange $\triangle$-s and blue $+$-s, respectively. The errors $\tilde{\mathcal{E}}$ calculated from (\ref{eqn:alg_ED_Eucl_norm}) between the evolution on a classical device (digital quantum simulator) and exact diagonalization results are shown in red $\times$-s (black $\circ$-s). Panel (a) shows the results for $m_{\text{lat}}/g=0$ for the parameter values set at the beginning of section~\ref{sect:num_an}. Panel (b): same as panel (a), but for $m_{\text{lat}}/g=10$. On the right side of the divider, panels (c), (d), (e) and (f) show the results for $m_{\text{lat}}/g=10$ with the following changes of parameter values: Panel (c): time-step $\delta t = 0.5$, $n_{\text{shots}} = 10^4$ shots for shot-induced noise (same as in panel (b)). Panel (d): $\delta t = 0.05$, $n_{\text{shots}} = 10^4$. Panel (e): $\delta t = 0.5$, without shot-induced noise. Panel (f): $\delta t = 0.05$, without shot-induced noise.}
    \label{fig:err_vs_iter}
\end{figure*}

In this section we consider the deviation $\mathcal{E}$ in (\ref{eqn:alg_Eucl_norm}) between two consecutive iterations of the algorithm, as well as the deviation $\tilde{\mathcal{E}}$ in (\ref{eqn:alg_ED_Eucl_norm}) of each individual iteration of the algorithm from the corresponding ED result. The results are shown in Fig.~\ref{fig:err_vs_iter}. Here, since we are able to analyze the system by ED, we are able to evaluate with certainty how our adiabatic method performs by comparing it with the true dependence of the energy levels on $l_0$.

Based on the results shown in Fig.~\ref{fig:err_vs_iter}, let us remark that an increase of the deviation $\mathcal{E}$ between consecutive iterations (orange and blue lines) does not determine whether $\tilde{\mathcal{E}}$ (red and black lines) increases or decreases. Instead, a greater value of $\mathcal{E}$ simply means that the absolute difference between consecutive iterations increased.

From the results in Fig.~\ref{fig:err_vs_iter}(a) for $m_{\text{lat}}/g=0$, we observe that both the evolution on a classical device as well as that on a digital quantum simulator approach the ED results as $T$ is increased (red and black lines, respectively). Furthermore, we observe that for $T$ approaching $500$ the difference between consecutive iterations both on a classical device and on a digital quantum simulator decreases. That is, the dependence of the levels on $l_0$ becomes less affected by the value of $T$. This further supports the validity of our results.

However, the results in Fig.~\ref{fig:err_vs_iter}(b) for $m_{\text{lat}}/g=10$ differ considerably. First, the deviation of the evolution on a classical device from the ED results (red line) does not decrease for $T$ approaching $500$. Second, the deviation of the evolution on a digital quantum simulator from the ED results (black line) increases. Moreover, clearly, the results obtained on the digital quantum simulator (black line) differ from those obtained on a classical device (red line). Consistently, the corresponding differences between two consecutive iterations on a classical device and on a digital quantum simulator (orange and blue lines, respectively) differ as well.

The reason for the mismatch between the results of the classical device and the digital quantum simulator lies in the difference of the corresponding methods. On the classical device, the solution for the time-evolution of a state is obtained by the integration of the set of ordinary differential equations determined by the system. On the digital quantum simulator the approach is different: This method utilizes Trotter expansion and is additionally affected by shot-induced noise. The fact that we observe the above discrepancies for the case of a level crossing and not in the case of a non-zero gap, suggests that these are gap-dependent.

To test whether this effect occurs because of the size of the Trotter step as well as due to the shot-induced noise, we perform supplementary simulations, where we vary the input parameters. The results are shown in Figs.~\ref{fig:err_vs_iter}(c) - \ref{fig:err_vs_iter}(f). Panel (c) shows the results of panel (b) on the domain $T \in [5, 50]$, this is the baseline for the comparison.

From panel (c) to panel (d) the change is that we decrease the Trotter step from $0.5$ to $0.05$. We observe that the digital quantum simulator (black line) now performs better for larger $T$. However, the result still differs from that of the classical device (red line). From panel (c) to panel (e) we remove the shot-induced noise from the simulation. Here, in fact, the digital quantum simulator performs worse for all $T$. Last, in panel (f) we decrease the time-step to $0.05$ and remove the shot-induced noise as well. Here we observe that the results of the classical device and of the digital quantum simulator match. This indicates that decreasing the size of the Trotter step as well as increasing the number of shots should improve the results on the digital quantum simulator. That is, we find evidence that our method can be successfully applied both on classical and on quantum devices.

Last, note that for $m_{\text{lat}}/g=10$ the deviation of the results on a classical device from the ED results (red line) exhibits local fluctuations over $T$ (see Figs.~\ref{fig:err_vs_iter}(b) - \ref{fig:err_vs_iter}(f)). This is due to the accumulation of numerical error: For these simulations, the adiabatic evolution of the levels closely recovers the ED results already for small values of $T$, with a corresponding not-normalized error of $\tilde{\epsilon} \tilde{\mathcal{E}} \lesssim 10^{-9}$ and the normalization factor of $\tilde{\epsilon} \lesssim 10^{-9}$.

\section{Discussion} \label{sect:disc}

\subsection{Comparison with the deformation method}

In this section we briefly review the adiabatic method of~\cite{sym14040809} and specify its similarities and differences to our method.

To find where the level crossing occurs, the idea of the method in~\cite{sym14040809}, which we refer to as the ``deformation method'', is the following: First, introduce into the Hamiltonian an additional term with a corresponding external parameter, such that the level crossings are avoided. Second, perform multiple simulations for decreasing values of the external parameter, determining the critical value of the other parameter originally responsible for the PT (i.e.\ the analog of $l_0^*$ for the Schwinger model). In~\cite{sym14040809}, for the considered example of the XY model, this corresponds to the midpoint of the step in the magnetization, which plays the role of the order parameter. Third, to find the PT point, extrapolate the obtained results into the limit of vanishing external parameter.

Before we highlight the differences of the two approaches, let us first comment on the common properties of the methods. The two approaches share a common advantage: Since both algorithms employ adiabatic preparation of states, it is reasonable to apply them to a problem where the spectrum can be easily found in a certain well-studied region of the parameter space, and more difficult to find in a less-investigated region that we would want to explore.

For a further common property, recall the discussion of section~\ref{subsect:adiab_ev_mult_st}: In order to ensure that an algorithm utilizing adiabatic state preparation on a quantum device performs correctly, we need to find indications for the convergence of the time-evolution when varying the number of shots $n_{\text{shots}}$, the time-step $\delta t$ and the total evolution time $T$. If the true phase structure is known, the convergence concerns recovering the ED results. If the phase structure is unknown, the convergence refers to the dependence of the energy levels on the parameter the variation of which is responsible for the PT. Let us again emphasize that while the manifestation of the latter does not necessarily imply that the former holds, this is the assumption that we make in~\ref{asmp:asympt_assumpt}. Note that the deformation method requires the same assumption about the asymptotic behavior of the energy levels.

In addition, for a general system, the deformation method faces the same difficulties when preparing the initial state. This would require an explicit state preparation, preparation by adiabatic state evolution, or by other means. We now consider the distinctions of the two approaches.

A particular distinction between the two methods becomes evident when the corresponding algorithms are executed on a quantum device. The deformation method requires adding an extra term to the Hamiltonian. First, clearly, this requires an understanding of what term to add. In general, this requires some analytic effort and knowledge of the system. Second, more importantly, adding any non-zero term to the Hamiltonian necessitates including additional corresponding gates into the circuit. On quantum hardware this entails amplifying the sources of errors occurring on the quantum device.

However, for a system with many level crossings, such as that studied in~\cite{sym14040809}, our method would require evolving a number of states much larger than only two, if the parameter range to be explored is broad. Below we expand on further differences of the two methods in detail, with the numerical results of section~\ref{sect:num_an} at hand.

In particular, it is instructive to compare our method with the adiabatic method of~\cite{sym14040809} based on the information that we are provided with for a given problem and the task at hand. The differences of the two methods are best characterized when categorized according to the amount and type of information that we a priori possess about the problem.

Here we identify three classes (or levels), depending on our knowledge of the system. To be specific, in each class we make the distinction between the two cases when the true behavior of the system is given by a PT and when there is a non-zero energy gap. For simplicity, we consider the case of a system of only two energy levels.

In general, if we know that there is a PT in the region of interest, we would like to find the parameter value for which it occurs. If we know that there is a non-zero gap, we would like to verify this and find its specific value. Finally, if we have no knowledge of the system, we would like to learn anything about its phase structure that we can.

\subsection{Dependence of the method on the knowledge about the system}

\subsubsection{Level 1: Perfect knowledge of the system}

At the ``easiest'' level, we possess perfect knowledge of the system, i.e.\ we know the phase structure for the parameter range that is of interest to us. Here, we would like to verify our understanding using the two adiabatic methods. In the case where we know that there is a minimal non-zero gap and we know its position and value, there is no need to deform the Hamiltonian. Therefore, both methods amount to performing the adiabatic state preparation for the two levels separately.

Now consider the case where we know that the gap is zero and that there is a crossing at a known value of the parameter. To recover the known results, following the deformation method, we add an auxiliary term to the Hamiltonian and, under the assumption~\ref{asmp:asympt_assumpt}, execute the algorithm for the ground state at least two times in order to perform the extrapolation. Applying our method, however, under the same assumption~\ref{asmp:asympt_assumpt}, we run the corresponding algorithm exactly two times - once for each of the two lowest levels.

Note that although both methods can find the critical value of the parameter responsible for the PT, the deformation method cannot recover the order of the PT. This can be achieved with our method for a PT of any order $n$. The corresponding evolution time $T$ will vary depending on the specific case at hand.

\subsubsection{Level 2: Limited knowledge of the system}

At this intermediate level, we have some knowledge of the phase diagram, for example from analytic considerations. Specifically, if we know that there is a non-zero gap, but we do not know its value and/or for which parameter value it is minimal, again, there is no need to deform the Hamiltonian and the strategy here is similar to that of \emph{Level 1} above.

If there is a level crossing at an unknown value of the parameter, the procedures for the two adiabatic methods are the same as for \emph{Level 1}.

\subsubsection{Level 3: No knowledge of the system}

For this level, which is the most challenging, we assume that we have no knowledge of the phase structure of the system. Specifically, consider the case when we don't know whether there is a PT or a minimal non-zero gap for some value of the parameter. Moreover, we have no knowledge of the corresponding critical parameter value.

Regarding the deformation method we note the following: If we deform the Hamiltonian and apply the corresponding algorithm, we may erroneously conclude that there is a PT where, in fact, there is none. Therefore, the deformation method is not applicable in this case. That is, the deformation method cannot be used when the phase structure of the model is not known.

However, this is where the procedure defined in~\ref{subsect:adiab_ev_mult_st} gains a clear advantage. Indeed, the application of this method remains viable for the present scenario: We evolve the two lowest levels and, under the assumption~\ref{asmp:asympt_assumpt}, we can argue to have found the underlying phase structure of the model for a specific parameter choice. If we find a level crossing, we can infer the order of the corresponding PT by considering the discontinuities in the derivatives of the energy of the ground state at the crossing point.

Although the scenario of no knowledge of the system is hypothetically possible, as discussed previously, analytical arguments, such as whether one expects a PT to occur or not, can greatly improve the predictive power of the methods. In the case that we truly know nothing about the system except for the Hamiltonian and the spectrum at one end of the parameter range, based on the adiabatic theorem of~\cite{Born1928}, our method still remains applicable. However, a physical model is usually constructed with some physical phenomenon or a defining characteristic in mind. This information may allow us to improve the interpretation of the results we obtain numerically.

\subsection{Limitations of the method}

Here we briefly recapitulate the limitations of our method. First, while our method can be advantageous if the two lowest energy levels are separated from the rest of the spectrum by a gap, more computational resources will be required if the higher excited levels cross with the lower two. In this case more states need to be evolved. If the underlying phase structure is a priori known to exhibit many level crossings involving multiple levels, such that each level can be identified with the ground state only for a short parameter range, the method of~\cite{sym14040809} may be more suitable for the task.

Second, similarly to the method of~\cite{sym14040809}, our method relies on the following assumption: If for increasing $T$ we observe convergence of the dependence of the individual energy levels on the parameter varied in the adiabatic evolution, we assume that this is the behavior of the levels for asymptotically large values of $T$.

Third, for particular problems where the sub-region of the Hilbert space that includes the ground state can be parametrized by a set of variables polynomial in the number of qubits, variational methods such as VQE can be computationally cheaper. However, if little is known about the ground state in the ``interesting'' parameter regime, and the variational ansatz circuit cannot be restricted to a sub-region of the Hilbert space that is polynomial in the number of qubits, or the optimization problem is difficult and the corresponding solution is unreliable, our method presents a viable approach.

\section{Conclusions and Outlook} \label{sect:concl}

In this work we introduced a new method for studying the phase structure of a quantum physical system on a quantum device. In particular, we applied it to investigate two distinct regimes: First, the occurrence of a first-order phase transition when varying a corresponding parameter of the model. Second, the absence of any transition where the ground state is separated from the rest of the spectrum by a non-zero energy gap.

The method is based on a direct application of the adiabatic theorem of~\cite{Born1928}: We individually evolve over a time $T$ the states from a sub-set of the lowest levels of the model with a time-dependent Hamiltonian, sweeping over the parameter range. Under the assumption that a convergence of the parameter-dependence of the energy levels for increasing times $T$ implies having reached their parameter-dependence for asymptotically large values of $T$, we recover the phase diagram of the model.

Our method is applicable if the phase structure of the model is unknown. Specifically, if there is a non-zero gap between the lowest-energy level and the rest of the spectrum, or a level crossing between the two lowest energy levels, we can recover it. Moreover, if the two lowest energy levels cross, we can infer the order of the corresponding phase transition from the discontinuities in the derivatives of the energy of the ground state at the crossing point.

We have successfully tested this method in application to the Schwinger model in the presence of a topological $\theta$-term. In this parameter regime, conventional Monte Carlo algorithms can no longer be applied reliably due to the manifestation of the sign problem. Specifically, we considered two parameter regions. First, high lattice mass-to-coupling ratio $m_{\text{lat}}/g$ with a corresponding first-order phase transition. Second, low $m_{\text{lat}}/g$, below the critical value, for which no transitions occur.

We compared our method with a different method~\cite{sym14040809} from the literature that also utilizes adiabatic state preparation. We have highlighted the similarities of the two methods and discussed their differences. In particular, our method does not require introducing additional terms into the Hamiltonian and, consequently, additional gates into the circuit that may lead to an increase in errors on quantum hardware. Moreover, our approach is applicable if nothing is known about the phase structure of the model.

In terms of future work, there are several promising directions. One of these is to consider various error mitigation schemes, such as, for example, zero-noise extrapolation~\cite{PhysRevX.7.021050, PhysRevLett.119.180509, 9259940, majumdar2023best} and global randomized error cancellation~\cite{PhysRevA.105.042408} in application to our method, and to compare them.


\begin{acknowledgments}

This work has received support from the French State managed by the National Research Agency under the France 2030 program with reference ANR-22-PNCQ-0002. We acknowledge the use of IBM Quantum services for this work. The views expressed are those of the authors, and do not reflect the official policy or position of IBM or the IBM Quantum team. O.K.\ would like to thank Stefan K\"{u}hn, Takis Angelides and Pranay Naredi for discussions on the Schwinger model and its implementation.

\end{acknowledgments}







\let\clearpage\relax

\end{document}